# Learning to decompose the modes in few-mode fibers with deep convolutional neural network

YI AN,[1] LIANGJIN HUANG,[1] JUN LI,[1,2] JINYONG LENG,[1] LIJIA YANG,[1] AND PU ZHOU[1,3]

[1]*College of Advanced Interdisciplinary Studies, National University of Defense Technology, Changsha 410073, China*
[2]*jun.johnson.li@gmail.com*
[3]*zhoupu203@163.com*

**Abstract:** We introduce a deep-learning technique to perform complete mode decomposition for few-mode optical fibers for the first time. Our goal is to learn a fast and accurate mapping from near-field beam patterns to the complete mode coefficients, including both modal amplitudes and phases. We train the convolutional neural network with simulated beam patterns and evaluate the network on both the simulated beam data and the real beam data. In simulated beam data testing, the correlation between the reconstructed and the ideal beam patterns can achieve 0.9993 and 0.995 for 3-mode case and 5-mode case, respectively. While in the real 3-mode beam data testing, the average correlation is 0.9912 and the mode decomposition can be potentially performed at 33 Hz frequency on a graphic processing unit, indicating real-time processing ability. The quantitative evaluations demonstrate the superiority of our deep learning–based approach.

## 1. Introduction

Recently, few-mode fibers (FMFs) have attracted much attention for both fundamental and applied research. Space division multiplexing based on FMFs is a promising way to overcome the anticipated capacity crunch of the single-mode fibers [1]. Larger mode area provided by FMFs helps to suppress the detrimental nonlinear effects and improve the damage threshold, which paves the way to higher power fiber lasers [2]. Furthermore, FMF is a perfect platform for experimental exploration on the complicated spatiotemporal soliton dynamics [3,4] and new nonlinear phenomena [5,6] in multi-mode fibers. With the rapid research progress of FMF, it is highly demanded to characterize the properties of the spatial modes emitting from the FMF, which is named as mode decomposition (MD) technique. With MD techniques, the amplitude and phase information of each eigenmode in the optical fiber can be estimated, providing the complete optical field and the beam properties associated with the field, e.g. wave front [7] and beam propagation factor [8]. Recent years, MD techniques have been widely used in many applications, such as optimizing fiber-to-fiber coupling [9], analyzing mode-resolved gain [10,11] or bend loss [12], diagnosing temporal mode instabilities [13,14], measuring mode transfer matrix [15,16] and realizing adaptive mode control [17,18].

In the past few years, various MD methods have been proposed with different techniques, such as spatially and spectrally resolved imaging [19], frequency domain cross-correlated imaging [20], ring-resonators [21], low coherence interferometry [22], correlation filter [23] and digital holography [24]. Although these methods can achieve accurate results, they require consuming post-data processing or intense experimental measurements. Besides these approaches, numerical computing-based MD methods have shown their equal accuracy without complex experimental operations [25–28].

Taking an intensity image of the beam as input, which is captured by a CCD camera, numerical computing-based MD methods usually use Gerchberg-Saxton [25], line-search [26], stochastic parallel gradient descent (SPGD) [27] or hybrid genetic global optimization algorithm [28] to find the optimal mode coefficients by minimizing the difference between the input beam profile and the reconstructed one. Gerchberg-Saxton algorithm requires heavy iterative calculation which is quite time-consuming [26]. The other algorithms, such as line-search [26], SPGD [27] and hybrid genetic global optimization [28] tend to search and update the mode coefficients gradually. However, the main drawback of these optimization methods is that they are very sensitive to the initial values and might trap into the local optima. Although real-time MD has been achieved by SPGD algorithm [27], only slowly varying beam can be processed with the unsatisfied processing frequency (~9 Hz).

Actually, real-time MD contributes a lot to dynamic monitoring of the spatial mode evolution of the FMF or optimizing the manufacturing of the mode-resolved devices. Recent progress on the successful applications of convolutional neural networks (CNN) in optics and photonics [29–32] has led many to ask whether a similar success is achievable in learning MD for the FMF or even multi-mode fiber. Comparing with the existing algorithms, the most attractive advantage of deep learning-based approach is that it can achieve excellent real-time performance with a trained neural network. To be specific, it only takes one forward pass (usually a few milliseconds) to perform MD and no initialization is required. We have noticed that deep learning technique has recently been demonstrated to be applicable for analyzing the modal power distribution in air-cladded silicon-on-insulator waveguide through simulations [33]. However, the modal phase information is ignored in their method. In practical condition, as the ground truth mode coefficients are unknown, the only reliable way to evaluate the accuracy of the inferred modal amplitudes and phases is to reconstruct the beam pattern and then measure its difference from the captured intensity image. Without phase information of each eigenmode, the beam pattern cannot be reconstructed for verifying the MD [33], which greatly hinders its practical application.

In this paper, we have developed an end-to-end deep neural network for very fast MD for FMF, learning how to map the input intensity images to the complete mode coefficient space. Unlike [33], our method can predict not only the modal weights but also relative phase between the higher order modes and the fundamental mode, leading to flexible applications in practical condition. Furthermore, we evaluate our method not only on simulated beam data but also real beam data, and the quantitative evaluations demonstrate the superiority of our deep learning strategy.

## 2. Methods

### 2.1. Modal decomposition basics

The propagating field in the optical fiber can be mathematically expressed as [34]

$$U(r,\varphi) = \sum_{n=1}^{N} \rho_n e^{i\theta_n} \psi_n(r,\varphi) \tag{1}$$

$$\sum_{n=1}^{N} \rho_n^2 = 1 \quad \theta_n \in [-\pi, \pi] \tag{2}$$

where $\psi_n(r,\varphi)$ is the electric field of the $n^{th}$ eigenmode in the fiber with modal weight $\rho_n^2$ and relative phase $\theta_n$. The eigenmodes could be described by linearly polarized (LP) modes based on weak-guidance approximation [34] and the number of them supported within the fiber depends on the fiber parameters. The purpose of the MD is to predict $\rho_n^2$ and $\theta_n$ from the near-field pattern.

One thing should be noted is that the sign of phase ambiguity exists with only one intensity image involved in the MD because the real and the conjugated fields cannot be distinguished in this case [26]. On the other hand, the ambiguity will not influence evaluation on the validity of the MD. Furthermore, providing accurate mode weights is sufficient in cases such as monitoring mode instability dynamics [15,16] and modal gain analysis [12,13].

*2.2. Deep-learning networks*

Convolutional neural network (CNN) has been proven to be highly effective for image processing [35–38], which contains different kinds of layers. Different layers in CNN have their particular functions. The convolution layer uses a set of filters to learn how to extract proper features from the input image for a specific task. The pooling layers progressively remove redundant information and reduce the size of feature maps to greatly decreases the computation and memory cost. The fully-connected layer converts the output of previous layers into the one-dimensional vector with a certain length, which represents a deep understanding of the whole image. The ReLU activation layer effectively removes negative values by setting them to zero to perform nonlinear activation.

We adopt the VGG-16 [35] model that has been pre-trained on the ImageNet data set and then the network is fine-tuned on our training data for MD. As shown in Fig. 1, this CNN model can be divided into 7 blocks, the details of the convolutional layers in each block are displayed in Table 1, including the number of convolutional layers (#conv), the size and the number of channels of the output feature map (#chan). The performances of max pooling and averaged pooling are investigated and we find the max pooling provides better performances. The reason might be that max pooling keeps the important features. Accordingly, a max pooling layer is added in the end of the first five blocks. The ReLU activation layer after each convolutional layer is omitted in Fig. 1 for better illustration. The last two blocks are two fully-connected layers. We change the filter size of the first convolutional layer from $3 \times 3 \times 3$ to $3 \times 3 \times 1$, as our input is a gray intensity image. The filter size and channels of the first fully-connected layer are set to $4 \times 4 \times 512$ and 1024 respectively. We modify the dimension of output vector of the last fully-connected layer according to the number of modes to ensure that the output vector size is equal to the label size. The Softmax layer of the origin VGG model is also replaced with a Sigmoid layer for our regression problem.

As shown in Fig. 1, the network learns to estimate mode weights and phases from a single gray image of the near-field beam intensity. During training, the input images are randomly generated based on the superposition of the theoretical eigenmodes calculated according to the known fiber parameters. The mode weights and phases are concatenated as a label vector. To be specific, if the number of eigenmodes supported within the fiber is $N$, then the mode weights $\{\rho_i^2 | i = 1, 2, \cdots N\}$ corresponding to all the eigenmodes and $N$-1 phase items $\{\theta_i | i = 2, 3, \cdots N\}$ denoting the relative phase differences between the high order modes and the fundamental mode are collected as a $2N$-1 dimensional label vector. We should notice here that, if the phases are directly used in the label vector, the network can't reach convergence because one image might have two labels due to the ambiguity of phases mentioned above, which may make CNN confused. To solve this ambiguity, we use cosine value to represent the real phases in label vector so that the consistency of training data can be guaranteed. The range of cosine value is linearly scaled from [-1,1] to [0,1], and the final sigmoid activation layer would ensure the validity of output predictions. We define the loss of our network as a mean-square error (MSE) between the output and the label vector, as shown in Eq. (3)

$$Loss = \frac{1}{M} \bullet \sum_{i=1}^{M} \sum_{j=1}^{2N-1} (y_o^{(i)}[j] - y_l^{(i)}[j])^2 \qquad (3)$$

where $y_o$ is the output vector, $y_l$ stands for the label vector and $M$ is the number of training samples.

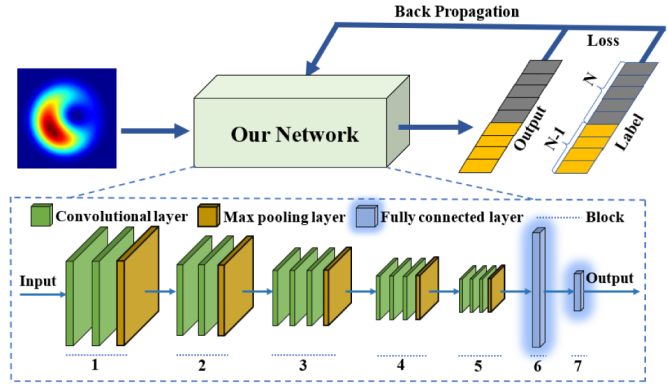

Fig. 1. Illustration of our network.

**Table 1. Details of our network**

| Network | Block | | | | | | |
|---|---|---|---|---|---|---|---|
| | 1 | 2 | 3 | 4 | 5 | 6 | 7 |
| #conv | 2 | 2 | 3 | 3 | 3 | 1 | 1 |
| size | 64 × 64 | 32 × 32 | 16 × 16 | 8 × 8 | 4 × 4 | 1 × 1 | 1 × 1 |
| #chan# | 64 | 128 | 256 | 512 | 512 | 1024 | 2N-1 |

In the back propagation stage of training procedure, the parameters of the network are updated iteratively through stochastic gradient descent (SGD) based on the MSE loss. The illustration of the testing procedure is shown in Fig. 2. In this step, we can obtain the predicted modal weights directly from the output vector of the network. For the relative phase, we collect all the possible combinations based on the estimated cosine values. The final predicted phase combination then can be determined from these candidates by searching the maximum of the correlation, which can be expressed as [26]

$$C = \left| \frac{\iint \Delta I_r(r,\varphi) \Delta I_m(r,\varphi) r dr d\varphi}{\sqrt{\iint \Delta I_r^2(r,\varphi) r dr d\varphi \iint \Delta I_m^2(r,\varphi) r dr d\varphi}} \right| \quad (4)$$

where $\Delta I_j(r,\varphi) = I_j(r,\varphi) - \bar{I}_j$ ($j = r, m$) and $\bar{I}_j$ is the corresponding mean value of measured beam intensity $I_m$ or reconstructed $I_r$. The value of $C$ indicates how similar is the reconstructed beam compared with the input. In the ideal case, the correlation has maximum 1 when the reconstructed pattern is the same with the input. Furthermore, there are two almost identical maximum correlations, corresponding to the real and the conjugated fields respectively.

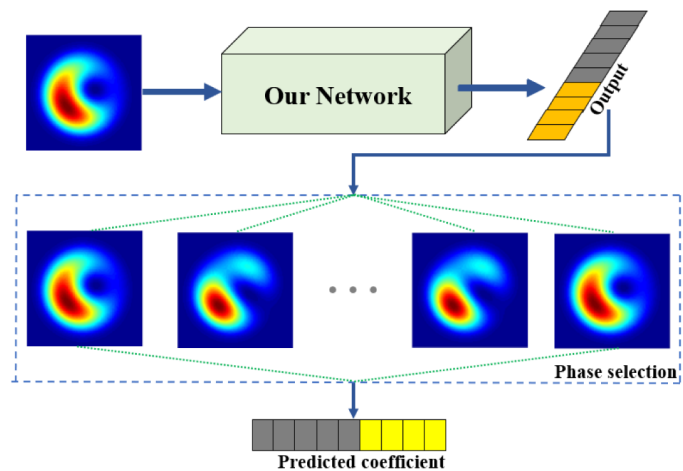

Fig. 2. Illustration of our testing procedure.

## 3. Results and discussion

All experiments reported in this paper are run on a desktop computer with an Intel Core i7-8700 CPU and GTX 1080 GPU. In each epoch, we randomly generate 100000 images on the fly with a resolution of 128 × 128 to train our CNN. The weights of the modified layers are initialized randomly and others are initialized by pre-trained VGG-16. We use min-batch SGD with batch size of 64 to accelerate computing speed, and the learning rate is set to 0.01 in the first 20 epochs and 0.001 in the following epochs. The network gets convergence after 30 epochs and the overall training time is 8 hours. We use both simulated and real captured beam patterns to evaluate the performance of our network.

### 3.1. Evaluating accuracy with simulated beam patterns

We take a step-index fiber with 25 μm core diameter and 0.08 NA working at 1064 nm wavelength as an example for simulation. The normalized frequency V of this fiber is about 5.91 so that it can support 10 modes, which can be arranged in order as $LP_{01}$, $LP_{11e}$, $LP_{11o}$, $LP_{21e}$, $LP_{21o}$, $LP_{02}$, $LP_{31e}$, $LP_{31o}$, $LP_{12e}$, and $LP_{12o}$ modes. Due to the degeneracy of the modes [34], there are 5 possible cases for the modes propagated in this fiber, which are the former 3, 5, 6, 8 and 10 modes respectively. Concentrating on 3-mode and 5-mode case, we train the CNN and test the network using 1000 beam patterns generated randomly. Testing patterns can be reconstructed with predicted coefficients of the CNN and this image reconstruction approach offers a visually and directly way to evaluate the MD accuracy of our scheme. We calculate the average correlation between the reconstructed patterns and ideal patterns of the testing samples after every training epoch and the results are shown in Fig. 3. Note that in 3-mode case the correlation increases to over 0.998 rapidly in only two epochs and finally close to 0.9993 after 30 training epochs while for 5-mode case the correlation converges to around 0.995 after 20 epochs. We also give some typical reconstructed and ideal patterns in Fig. 4 with the network trained for 30 epochs. The residual intensity patterns which can be achieved by $\Delta I = |I_m - I_r|$ and the corresponding correlation value are also provided in Fig. 4. It is found that the reconstructed beam patterns are of great similarity to the input ideal beam patterns and the residual intensity patterns are quite insignificant, which indicates the high accuracy of our scheme.

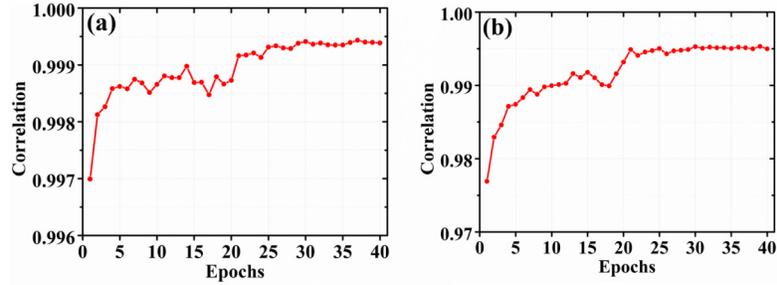

Fig. 3. Averaged correlation as a function of epochs for two cases. (a) 3-mode case; (b) 5-mode case.

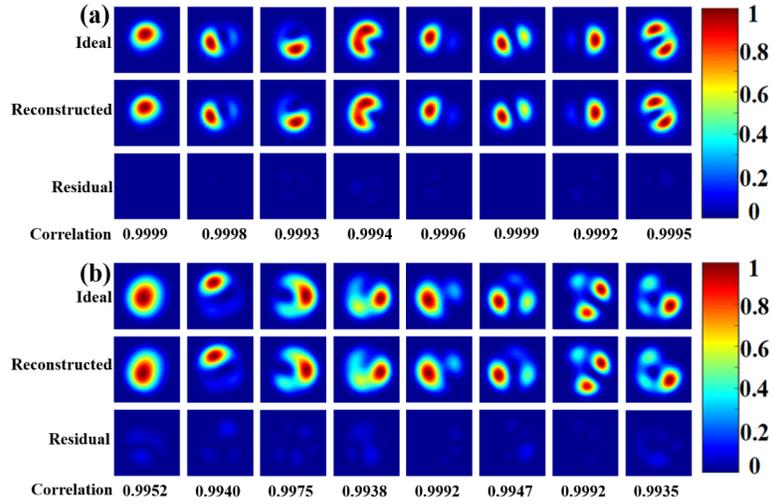

Fig. 4. Typical MD examples for two cases. (a) 3-mode case; (b) 5-mode case.

Because the mode coefficients of the simulated samples are known, here we also utilize a statistical and numerical method to analyze the precision. We calculate the absolute error between the predicted and the true coefficients for the convergent network which has been trained for 30 epochs. We define weights error $\Delta\rho^2$ and phase error $\Delta\theta$ as $|\rho_p^2 - \rho_t^2|$ and $\||\theta_p| - |\theta_t|\|/2\pi$ respectively, where $p$ and $t$ denote predicted and true coefficient. The definition of phase error is based on the fact that the predicted phase might be the opposite of the true value, which cannot be avoided with only one beam profile involved in MD [26]. The average errors of 1000 testing samples for modal weights and relative phase of both separate and whole modes are listed in Table 2 and Table 3 respectively, utilizing a centesimal system for better comparison. It is can be found in Table 2 that the averaged prediction error of modal weights is about 0.5% and 1% for 3-mode and 5-mode case respectively while the average modal phase prediction error in Table 3 is around 0.7% or 1.3% in two cases. According to the correlation value and the calculated error, we can find the accuracy for 5-mode case is slightly lower than 3-mode.

Table 2. Averaged error of modal weights

| | $\overline{\Delta\rho_{01}^2}$ | $\overline{\Delta\rho_{11e}^2}$ | $\overline{\Delta\rho_{11o}^2}$ | $\overline{\Delta\rho_{21e}^2}$ | $\overline{\Delta\rho_{21o}^2}$ | $\overline{\Delta\rho^2}$ |
|---|---|---|---|---|---|---|
| 3-mode case | 0.55% | 0.50% | 0.45% | - | - | 0.50% |
| 5-mode case | 0.82% | 1.27% | 1.37% | 0.84% | 0.87% | 1.03% |

Table 3. Averaged error of modal phase

|  | $\overline{\Delta\theta_{11e}}$ | $\overline{\Delta\theta_{11o}}$ | $\overline{\Delta\theta_{21e}}$ | $\overline{\Delta\theta_{21o}}$ | $\overline{\Delta\theta}$ |
|---|---|---|---|---|---|
| 3-mode case | 0.71% | 0.69% | - | - | 0.70% |
| 5-mode case | 1.17% | 1.24% | 1.45% | 1.46% | 1.33% |

Table 4. Consuming time for the CNN

|  | Total Time | | Averaged Time | |
|---|---|---|---|---|
|  | CPU | GPU | CPU | GPU |
| 3-mode case | 71s | 9s | 71 ms | 9ms |
| 5-mode case | 76 s | 14 s | 76 ms | 14 ms |

　　To measure the cost performance of MD, we run our method on 1000 testing images one by one using the trained network, solely using CPU or GPU respectively. The total and averaged time is reported in Table 4. We can see that the performance is significantly improved by using GPU due to its powerful parallelism calculation ability. Noted one beam profile only takes ~10 ms to perform MD, indicating feasibility of excellent real-time MD. The video of the processing results through the network with GPU is illustrated in Visualization 1, with slow 10 × illustrated in Visualization 2. Additionally, the video corresponds to the results shown in Fig. 4(a).

　　We also train the network for 6-mode, 8-mode and 10-mode cases to get convergence. The relation between the mode number and correlation as well as average error are shown in Figs. 5(a) and 5(b), respectively. We can find that the correlation decreases and the average prediction error of both modal weights and phase items get larger when the mode number increases. For 10-mode case, the averaged weights and phase error reaches about 2.3% and 4.9% respectively while the correlation drops to about 0.95. The reason might be that when the modes are augmented, the number of similar near-field beam patterns with different weights and phase combinations increases, which is another kind of ambiguity and enlarges the complexity [26]. Patterns with higher resolution may help mitigate this ambiguity as the details of the similar patterns can be observed. We have utilized 5-mode patterns with a resolution of 224 × 224 to train and test CNN, and the averaged prediction error descend from 1.03% and 1.33% to 0.93% and 1.26% for weights and phase respectively. However, the consuming time of training CNN with such patterns for 30 epochs increases to 24 hours, three times more than the cost of the 128 × 128 resolution. Based on these calculated results, we conclude that our scheme is robust to achieve an accurate MD result for the beams consisting of six or fewer modes of FMF. The decomposition for multi-mode beams consisting of eight or more modes will need patterns with higher resolution and the accuracy will possibly decrease because of the finite image resolution and the highly increasing ambiguous patterns. In order to enhance the MD accuracy of these multi-mode beams, we can take far-field beam profile as additional input, because the similar near-field beams with different mode coefficients have totally different far-field beam patterns. With both near-field and far-field beam patterns, the mode coefficients can be uniquely determined and almost no ambiguity exists [26]. Combining our scheme with SPGD [27] can also increase the accuracy for multi-mode beams. That is, the mode coefficients obtained from CNN can be set as the initial values of SPGD algorithm and then more accurate MD result can be achieved after iterations of SPGD.

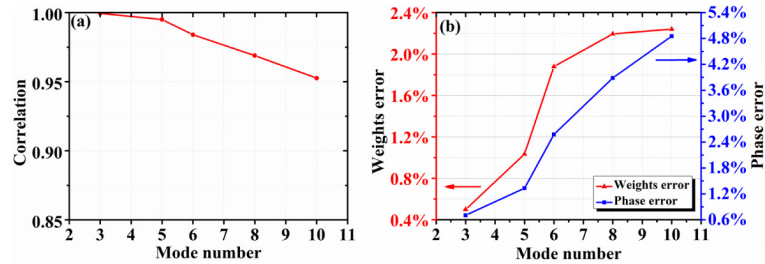

Fig. 5. The relation between the mode number and (a) correlation; (b) weights and phase error.

Taking 5-mode case as an example, the robustness of CNN is investigated by adding noise to the input pattern. The ground truth is the ideal clean beam pattern. For the generation of noisy patterns, every pixel of ground truth is multiplied by a factor, which equals to $1 + N(0,1) \cdot \sigma$. Here $N(0,1)$ is the standard normal distribution and $\sigma$ is defined as noise intensity [33]. We generated 1000 testing samples under different noise intensity and fed them into the trained CNN. A few input noisy patterns with different $\sigma$ value and the corresponding reconstructed ones are shown in Fig. 6(a). The averaged correlation between the reconstructed pattern and the ground truth is plotted in Fig. 6(b). Notice that the correlation value is still over 0.99 even when $\sigma$ reaches 0.32, showing a high anti-noise ability of our CNN. We also calculated the averaged weights and phase error for different noise levels, as shown in Fig. 6(c). It is found that even when the noise intensity increases to 0.32, the weights and phase error are still lower than 1.8%, which further proves the high anti-noise ability of our trained CNN.

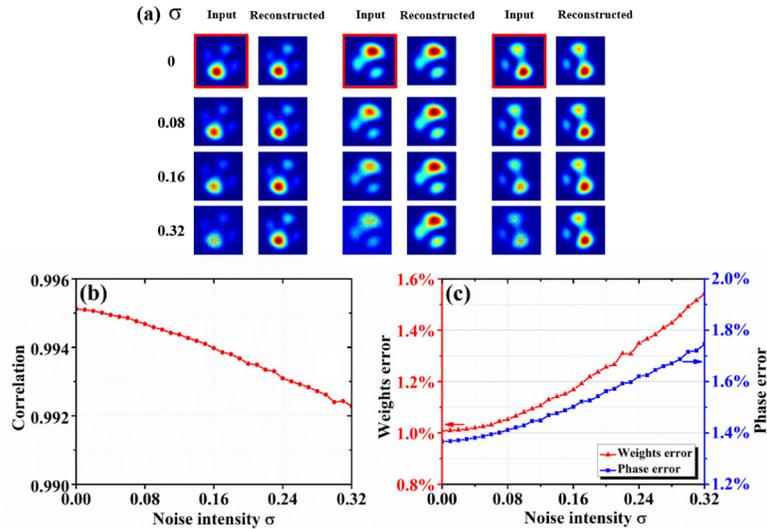

Fig. 6. The performance of CNN for noisy input patterns. (a) input patterns and reconstructed ones under different noise intensity levels. The pattern in the red rectangle is the ground truth. (b) averaged correlation between the reconstructed pattern and ground truth under different noise intensity. (c) averaged weights and phase prediction errors under different noise intensity.

### 3.2. Processing recorded frames of CCD

The experimental setup to acquire real beam patterns emitting from the optical fiber is shown in Fig. 7. We adopt a pig-tailed narrow linewidth laser diode at 1073 nm as the laser source. The delivery fiber is single-mode. Then the optical beam is coupled into the FMF with the core diameter of 8.2 μm and the NA of 0.14. The V-value of this FMF is 3.36 at 1073 nm, thus only supporting the $LP_{01}$, $LP_{11e}$ and $LP_{11o}$ mode. The output beam from the end facet of the fiber is imaged on the CCD camera (SP620U) through a 4-*f* imaging system consisting of 2 lenses whose focal lengths are 8 mm and 400 mm respectively. Accordingly, the magnification factor of this lens set is approximately 50. A polarization beam splitter is located between the lenses to select only one polarization component of the beam and the half-wave plate is to change the polarization. To obtain varying beam patterns, we rotate the half-wave plate to change the polarization of the emitting light and 16 constant frames are recorded automatically by CCD. Then we train the CNN by simulated beam patterns according to the parameters of the fiber and the optical system for 30 epochs and the network gets convergence. Then this convergent network is utilized to perform the MD on the experimental data. Because the parameters of CNN have been fixed after the training procedure, we can directly obtain the predicted output vector through only one forward pass of network. Since the resolution of the original captured images is 768 × 1024, we crop and resize these 16 frames to 128 × 128 to place the pattern in the center of image, and then pass them to our network to predict the final mode coefficients. The total cost time for processing these 16 frames is 480 ms, which means the MD frequency of our approach can reach ~33 Hz, showing much better real-time performance than SPGD (~9 Hz) [27]. The video of the processing results through the network with GPU is illustrated in Visualization 3, with slow 3 × illustrated in Visualization 4. Note that the cost is higher than the averaged time in Table 4, since processing the experiment raw data into the required size and format for CNN takes extra time. The performance can be further improved by optimizing the collecting and pre-processing procedure.

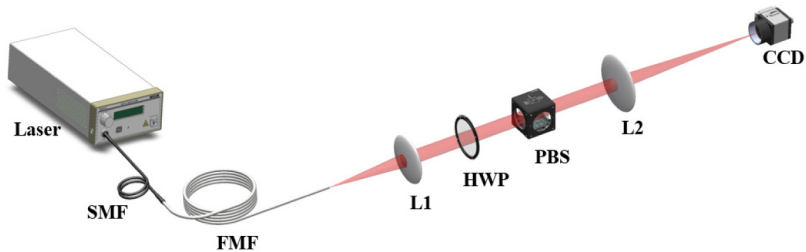

Fig. 7. Scheme of the experimental setup. SMF, single mode fiber; FMF, few-mode fiber; L, lens; HWP, half-wave plate; PBS, polarization beam splitter.

Figure 8 illustrates the evolution of the correlation between the measured and the reconstructed intensity pattern of every frame from one group of experimental data. Most of them reach over 99% and their mean value is 99.12%, indicating high accuracy of the decomposition by CNN. This can also be verified by not only the high agreement between the measured and the corresponding reconstructed patterns but also the negligible residual patterns as shown in the insets of Fig. 8. Additionally, the noises of captured images may bring negative effects to the accuracy of MD so that our correlation is a bit smaller than simulation cases.

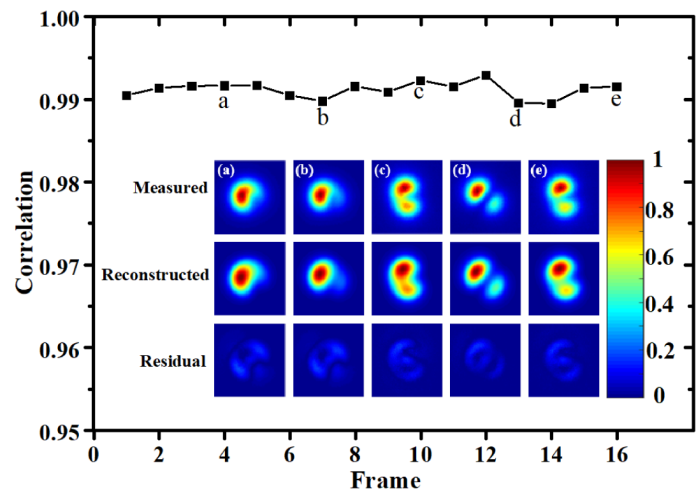

Fig. 8. The correlation between the measured and the reconstructed intensity pattern of every frame. The insets depict measured (up), reconstructed (middle) and residual(down) intensity of the frame denoted with letters.

## 4. Conclusion

We proposed a deep learning-based complete MD technique for FMF with potential real-time performance and high accuracy. By utilizing our CNN, the MD speed can be 30 ms per frame on GPU. The quantitative evaluations on both simulated and real beam patterns demonstrate the superiority of our method. With the work presented, we have only made a first step towards deep learning-based MD. In the future, we will further investigate the deep learning-based MD strategy for employing both near- and far- field beam patterns and for multi-mode fibers.

## Funding

National Natural Science Foundation of China (NSFC) (61605246, 61805280); Research Grants from College of Advanced Interdisciplinary Studies, National University of Defense Technology (JC18-07).

## Acknowledgments

We thank Dr. Jian Wu for providing help in the manuscript writing.